\documentclass[aps,prb,reprint,superscriptaddress,showpacs]{revtex4-1}

\usepackage[utf8]{inputenc}  
\usepackage{graphicx}
\usepackage{hyperref}
\usepackage{color}
\begin{document}


\title{Anharmonic suppression of Charge density wave in 2H-NbS$_2$}



\author{M. Leroux}
\affiliation{Institut N\'EEL CNRS/UJF - Grenoble, France}

\author{M. Le~Tacon}
\affiliation{Max-Planck-Institut~f\"{u}r~Festk\"{o}rperforschung, Heisenbergstr.~1, D-70569 Stuttgart, Germany}

\author{M. Calandra}
\affiliation{CNRS, Institut de Min\'eralogie et de Physique des Milieux Condens\'es, case 115, 4 place Jussieu, 75252 Paris Cedex 05, France}

\author{L. Cario}
\affiliation{Institut des Mat\'eriaux Jean Rouxel (IMN), Universit\'e de Nantes, CNRS, 2 rue de la Houssini\`ere, BP3229, 44322 Nantes, France}

\author{M-A. M\'easson}
\affiliation{Laboratoire Mat\'eriaux et Ph\'enom\`enes Quantiques, UMR 7162 CNRS, Universit\'e Paris Diderot Paris 7, B\^at. Condorcet, 75205 Paris
Cedex 13, France}

\author{P. Diener}
\affiliation{SRON Netherlands Institute for Space Research - Utrecht, Netherlands}

\author{E. Borrissenko}
\affiliation{European Synchrotron Radiation Facility - Grenoble, France}

\author{A. Bosak}
\affiliation{European Synchrotron Radiation Facility - Grenoble, France}

\author{P. Rodi\`ere}
\email[Corresponding author: ]{pierre.rodiere@grenoble.cnrs.fr}
\affiliation{Institut N\'EEL CNRS/UJF - Grenoble, France}

\date{\today}

\begin{abstract}
The temperature dependence of the phonon spectrum in the superconducting transition metal dichalcogenide 2H-NbS$_2$ is measured by diffuse and inelastic x-ray scattering. A deep, wide and strongly temperature dependent softening, of the two lowest energy longitudinal phonons bands, appears along the $\mathrm{\Gamma M}$ symmetry line in reciprocal space. In sharp contrast to the iso-electronic compounds 2H-NbSe$_2$, the soft phonons energies are finite, even at very low temperature, and no charge density wave instability occurs, in disagreement with harmonic ab-initio calculations. We show that 2H-NbS$_2$ is at the verge of the charge density wave transition and its occurrence is only suppressed by the large anharmonic effects. Moreover, the anharmonicity and the electron phonon coupling both show a strong in-plane anisotropy.
\end{abstract}

\maketitle

\section{Introduction}
In transition metal dichalcogenides (TMD) the interplay between strong electron-electron and electron-phonon interactions produces a wide variety of  instabilities ranging from charge density wave (CDW), as in TaSe$_2$, TaS$_2$, NbSe$_2$ and TiSe$_2$~\cite{Morosan2006,Cercellier2007,KusmartsevaPRL2009}, to Mott-insulating states in TaS$_2$~\cite{Sipos08}. The most eminent representative of this class of materials is 2H-NbSe$_2$ in which a CDW ($T_\mathrm{CDW}=33\,$K) coexists with superconductivity ($T_\mathrm{C}=7.1\,$K) with a strongly anisotropic superconducting gap. The debate about the origin of the CDW instability in this system has recently been revived. Several independent studies tried to locate the CDW gap using Angle Resolved Photo Emission Spectroscopy (ARPES)~\cite{RossnagelARPES2001,VallaPRL2004,Kiss07,Shen_ARPES_NbSe2,Borisenko09}. These studies
however found either Fermi surface nesting inconsistent with the CDW wave vector or no Fermi surface nesting at all. As such these studies were not conclusive concerning the origin of the charge density wave instability in TMD.
On the other hand, theoretical calculations~\cite{Mazin2006, Calandra_PRB2009} based on density functional theory (DFT) reproduce the experimental CDW state, despite a real part of the bare electronic susceptibility having only a broad maximum rather than a sharp divergence. In contradiction with a pure nesting scenario, the authors conclude that the main driving force for the CDW transition is the electron-phonon coupling (EPC). The accuracy of these theoretical calculations was recently validated against experiment by direct measurements of the phonon dispersion using inelastic x-ray scattering (IXS)~\cite{Weber_NbSe2}.

In most metallic TMD, superconductivity coexists with a charge ordered state or at least, it occurs in proximity of such a state. For many years it has been thought
that the occurrence of a CDW state was mandatory for superconductivity. However there exists at least one case that disagrees with this picture, namely 2H-NbS$_2$. This compound is isoelectronic and isostructural to 2H-NbSe$_2$, and has a similar superconducting transition temperature  ( $T_\mathrm{C}= 6\,$K).  However no CDW order occurs in 2H-NbS$_2$~\cite{guillamonPRL08,TMD_Naito81}, which does not fit really well with the variety of electronic instabilities found in the TMD family.

In this work we measure the phonon dispersion and linewidth of 2H-NbS$_2$, using thermal diffuse scattering (TDS) and IXS. Moreover, we perform first-principles linear-response calculations of the harmonic phonon dispersion. Surprisingly we find that while theoretical calculations do predict the occurrence of a CDW state, in experiments only a wide and marked softening of the low energy phonon branches is present, meaning that no CDW state occurs down to 2K. We demonstrate that this disagreement is due to the occurrence of large anharmonic effects that prevent CDW formation and stabilize the lattice. Our work shows that 2H-NbS$_2$ is, actually, almost as close to the CDW instability as all other metallic TMD and points out the relevance of anharmonic effects to describe the competition between CDW and superconductivity in TMD. Moreover, we show that the EPC is strongly anisotropic, giving a natural explanation to the superconducting gap anisotropy. 

\begin{figure}
  \includegraphics[width=8cm]{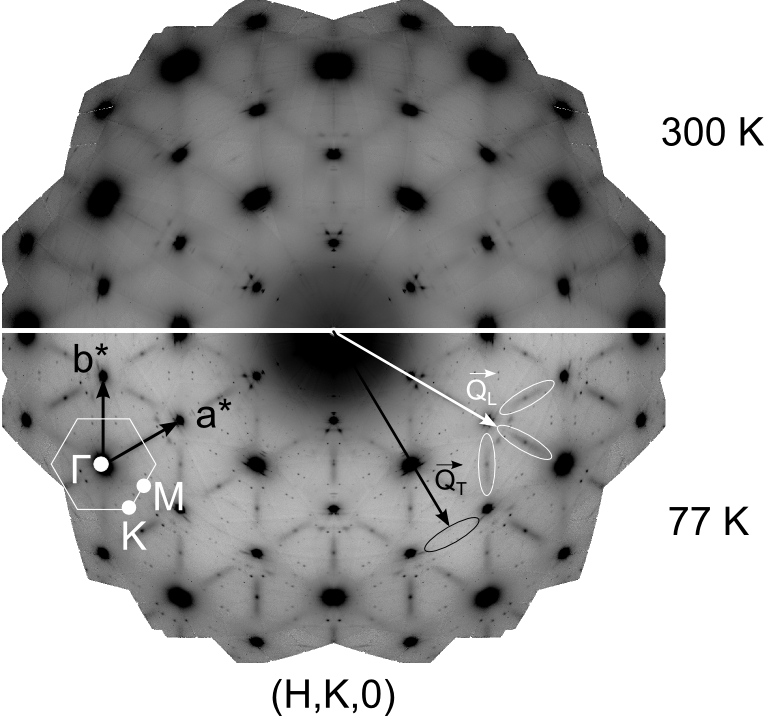}
  \caption{Reconstruction of the x-ray diffraction in the (H,K,0) plane of 2H-NbS$_2$ at 300 K and 77 K, from a 3D mapping of thermal diffuse scattering. A symmetrization has been applied. Diffuse intensity is present only along $\mathrm{\Gamma M}$ and, more precisely, only where scattering geometry selects longitudinal phonons (white ellipses). No diffuse features are visible for transverse geometry (black ellipse). At 77K small extra spots are visible.}
  \label{fig:diffuse}
\end{figure}

\section{Experimental details}

High quality single crystals of 2H-NbS$_2$ were prepared by iodine vapor transport with a large excess of sulphur in a sealed quartz tube. Powder diffraction on several batches gave a predominance of 99\% for 2H polytype ($P6_3/mmc$ space group), and 1\% for 3R. The polytype was then fully checked by x-ray diffraction for each selected single crystal. Refined lattice parameters at room temperature are $a=b=3.33\,$\AA,~$c = 11.95\,$\AA. The superconducting properties of the main sample used for this study were published elsewhere~\cite{Kacmarcik2010,Diener2011}.

TDS experiments~\cite{Bosak_TDS} at 77 K and room temperature were performed using a PILATUS 6M detector and a wavelength $\lambda=0.6966\,$\AA~(Fig.~\ref{fig:diffuse}).~In the ({\bf a},{\bf b}) plane, the Bragg peaks are sharp: full width of the (1,1,0) reflection is 0.13$\,^\circ$, evidencing a good structural order. Meanwhile along the c-axis, analysis of the diffraction pattern revealed a correct 2H stacking over 3 unit cells on average, in agreement with a previous study~\cite{Katzke2002}. Between the Bragg spots in the ({\bf a},{\bf b}) plane, diffuse lines can be detected at room temperature, and their intensities increase at low temperature. These lines are present only along the $\mathrm{\Gamma M}$ direction and, more precisely, only where scattering geometry selects longitudinal phonons (circled in white in Fig.~\ref{fig:diffuse}). No such diffuse features are visible for transverse polarization (circled in black in Fig.~\ref{fig:diffuse}). This indicates that an anomaly occurs for longitudinal phonons along $\mathrm{\Gamma M}$. In addition, at low temperature, we observe the apparition of tiny spots at wavevectors ${\bf q} = (0.38,0.16,0)$ and at~the~M~point~${\bf q} = (0.5,0,0)$.

\begin{figure}
\includegraphics[width=8cm]{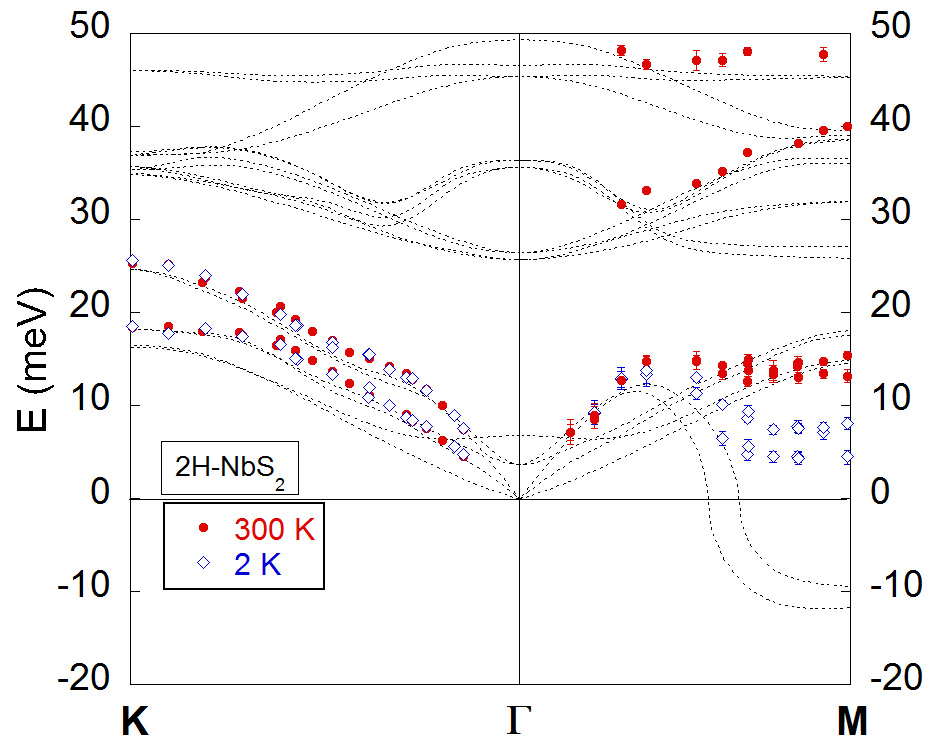}
\caption{(Color online) Experimental phonon spectrum of 2H-NbS$_2$ at room temperature (full red circles) and 2K (open blue diamond), compared to ab initio, zero temperature, harmonic phonons calculation (dashed lines). The two soft phonon branches along $\mathrm{\Gamma M}$ are nearly degenerate at room temperature, and slightly split away as temperature decreases. At the exception of the scans along $\mathrm{\Gamma M}$ at room temperature, we limited ourselves to energies below 30 meV.
}
\label{fig:dispersion}
\end{figure}

\begin{figure}
\includegraphics[width=7cm]{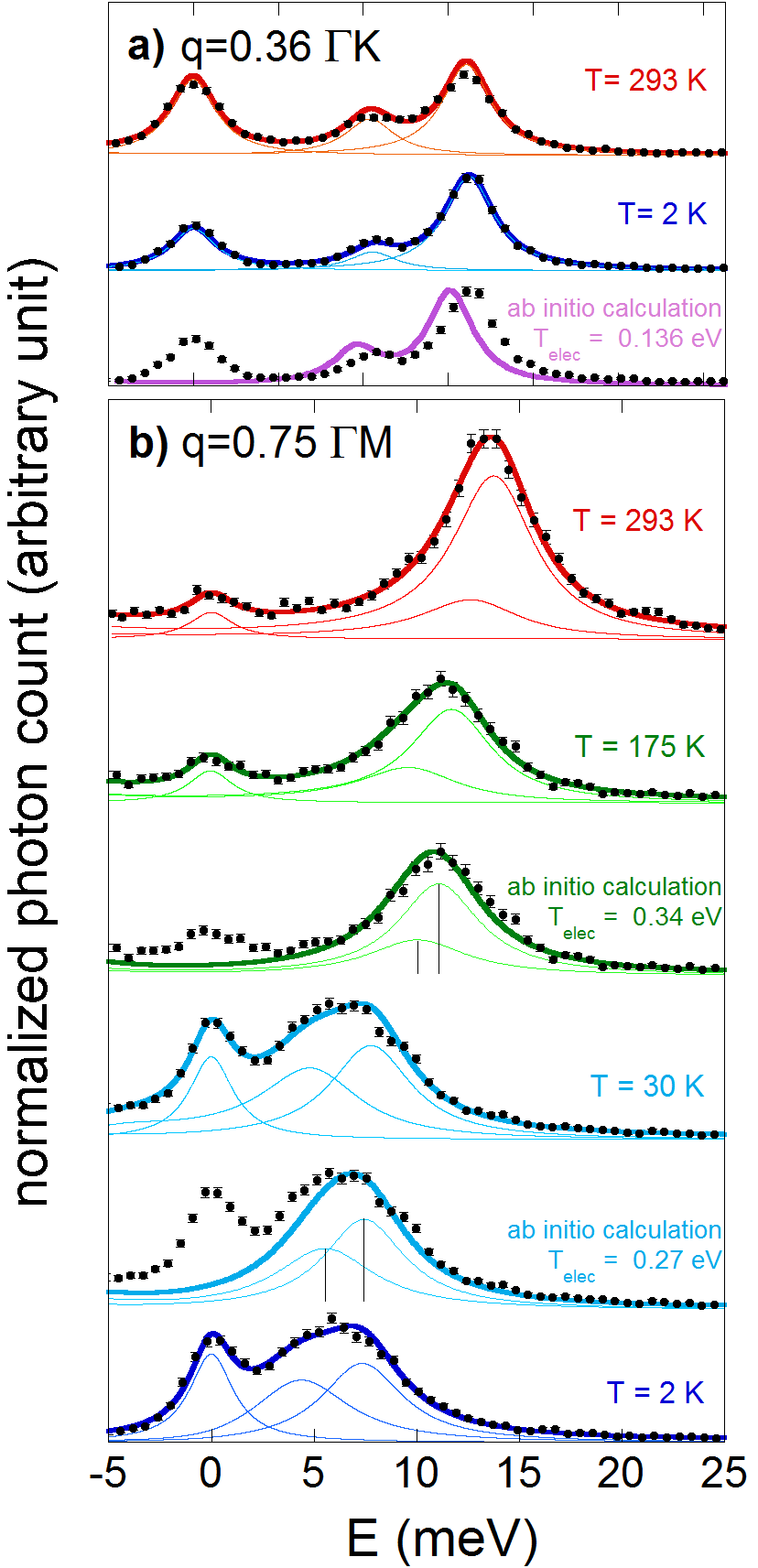}
\caption{(Color online) a) IXS scans at (H,K,L)=(3.184,-0.072,0) fitted by the convolution of a DHO (with Bose factor) and the Lorentzian experimental resolution. Ab initio phonon calculations, convoluted by the experimental resolution, are in good agreement with experimental results.\\b) IXS scans at (H,K,L)=(3.375,0,0) for different temperatures. For clarity, scans are shifted up. In the superimposed fits, the linewidths and amplitude ratio of the two phonons are fixed according to the ab-initio calculations (see text). Ab initio phonon calculations, convoluted by the experimental resolution, are also in good agreement with experimental results when using higher $T_\mathrm{elec}$.}
  \label{fig:scans_allT}
\end{figure}

To get further insight regarding these diffuse features, we performed IXS measurements with a photon energy of 17.794 keV using the (9,9,9) reflection on the high-resolution silicon backscattering monochromator. Corresponding instrumental energy and momentum resolutions were 2.6 meV FWHM (determined from the least-square fit of the Lorentzian curve of the elastic peak), $0.017\,$\AA$^{-1}$ in the (H,K,0) plane and $0.04\,$\AA$^{-1}$ in the (0,0,L) direction, respectively. The x-ray beam was focused down to $200 \times 60 \: \mu	m$ (width $\times$ height) on a single crystal of 2H-NbS$_2$ of $450^2 \times 100\,\mathrm{\mu m}$ (a $\times$ b $\times$ c). We found the tiny spots to be very weak elastic peaks, considering the high electronic susceptibility at ${\bf q} = (0.5,0,0)$, these may be traces of a charge density modulation induced by local strain or of Friedel's oscillations. Meanwhile, the diffuse features were due to a strong phonon softening. This can be evidenced when mapping the phonon dispersion (see Fig.~\ref{fig:dispersion}) along the $\mathrm{\Gamma K}$ and $\mathrm{\Gamma M}$ line at fourteen different temperatures between 300 K and 2 K (pumped $^4$He cryostat). We performed our measurements close to the (3,0,0) Bragg reflection, where we should observe both supposedly soft modes on LO and LA branches, according to structure factor calculations. The presence of these two modes is confirmed experimentally by the asymmetric profile of the experimental spectra in Fig.~\ref{fig:scans_allT} (b).
Harmonic phonon frequencies were calculated in the framework of DFT in the linear response~\cite{QE-2009}. We use the local density approximation. The dynamical matrix were calculated on a $10\times 10\times 1$ phonon-momentum grid in the Brillouin zone and Fourier interpolated throughout the Brillouin zone. The electronic integration was performed using $20\times 20\times 6$ electron-momentum grid and an Hermite-Gaussian smearing (electronic temperature) $T_{\rm elec}=0.136\,\mathrm{eV}$. The EPC contribution to the phonon linewidth (a quantity that does not depend on real or imaginary nature of the phonon frequency) has also been calculated following Ref.~\cite{Allen}. Experimental spectra were fitted using standard~\cite{Fok} damped harmonic oscillator (DHO) functions for the phonons with the  Bose factor included. The linewidths were constrained to calculated electron-phonon ones (see Tab.~\ref{tab:HWHM}), convoluted by the quasi-Lorentzian experimental resolution function. This procedure yields excellent results as seen in Fig.~\ref{fig:scans_allT}, and the resulting experimental dispersion of the phonons is plotted in Fig.~\ref{fig:dispersion}, together with the calculated harmonic phonon frequencies.

\section{Results and discussion}

Along the $\mathrm{\Gamma K}$ direction the low energy phonon dispersion is in excellent agreement with harmonic first-principles calculations, as also shown by the scans in Fig.~\ref{fig:scans_allT} (a). On the contrary along $\mathrm{\Gamma M}$ and close to the M point, harmonic calculations do show imaginary phonon frequencies for the lowest energy phonon modes, meaning that the high temperature structure becomes unstable and that a CDW with ordering vector close to the 0.5~$\bf a$* occurs. Experiments find a completely different picture. Namely, a marked softening of these branches occurs but the system remains dynamically stable with no indication of CDW instabilities.

\begin{figure}
  \includegraphics[width=8cm]{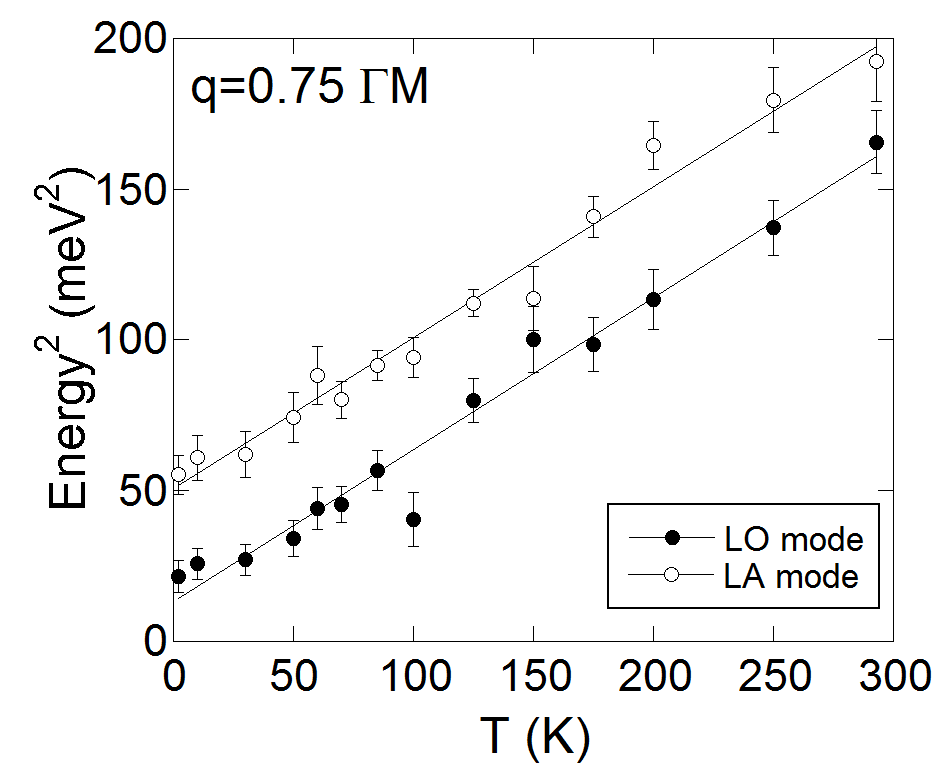}
  \caption{Experimental phonon energies, squared, from our fitting procedure with two phonons (see text), as a function of temperature at (h,k,l)=(3.375,0,0). Energies are compatible with the mean-field theory which predicts a power law $T^\frac{1}{2}$. Lines are guides to the eyes.}
  \label{fig:softmodeTdependence}
\end{figure}

More insight can be obtained by analyzing the behaviour of the measured phonon frequencies as a function of temperature. Experimentally, along the $\mathrm{\Gamma K}$ direction (see Fig.~\ref{fig:dispersion}) the phonon frequencies are essentially temperature independent. On the contrary, along $\mathrm{\Gamma M}$, and in particular close to the M point, the low energy phonon modes are strongly temperature dependent. This is better seen in Fig.~\ref{fig:softmodeTdependence} where the lowest energy phonon frequencies at 0.75 $\mathrm{\Gamma M}$ are plotted as a function of temperature. The phonon frequencies are more than doubled in the $2-300\,\mathrm{K}$ temperature range, a behaviour that can only be explained by invoking strong anharmonic effects~\cite{Calandra_PhysC}, as discussed hereafter. 

In solids, anharmonicity has two main effects. Firstly, it is the origin of the thermal expansion in solids. However this contribution is negligible here, as the lattice parameters barely change from zero to room temperature (a$(300\,\mathrm{K})=3.3295\,$\AA, a$(2\,\mathrm{K})=3.3230\,$\AA). Secondly, when the lattice parameters are fixed, phonon-phonon scattering results in a temperature dependent phonon-frequency shift to higher energies and in a temperature dependent enhancement of the phonon-linewidth~\cite{Calandra_PhysC}.  These contributions add to the normal electron-phonon contributions  that are, on the contrary, temperature independent. Thus, a strong temperature dependence of the phonon frequencies is a fingerprint of large anharmonic effects. In fact, we find here that the square of the two phonons frequencies at the softest {\bf q} vector increases linearly with temperature (Fig.~\ref{fig:softmodeTdependence}). This power law dependence of the frequency, also observed in 2H-NbSe$_2$~\cite{Weber_NbSe2}, is typical for a soft mode. As temperature is increased, the unstable harmonic oscillator is stabilized by higher order anharmonic potential, induced by the phonon-phonon interactions~\cite{Scott_RMP1974}. The critical exponent $1/2$ indicates that the phonon-phonon interaction can be described in a mean-field approach
, \textit{i.e.} considering the soft mode in equilibrium with the average field induced by all other phonons~\cite{Scott_RMP1974}.


First principles evaluation of third and fourth order perturbative coefficients in phonon-phonon scattering cannot be easily done here as the harmonic solution has imaginary phonon frequencies as seen in Fig.~\ref{fig:dispersion}. As such, the anharmonic correction is as big as the bare harmonic phonon frequency itself. Non perturbative approaches which include anharmonic effects~\cite{Errea_PRL2011} are required. These approaches are however very time consuming and unfeasible for the case of 2H-NbS$_2$ with $6$ atoms per cell.
In order to interpret the IXS spectra along $\mathrm{\Gamma M}$, we artificially smear out the sharp Fermi surface with an electronic temperature
a technique also used in Ref.~\cite{Weber_NbSe2}.

We have calculated the contribution of the electron-phonon coupling to the linewidths of the phonon modes, along the two high symmetry lines $\Gamma M$ and $\Gamma \mathrm{K}$. The values of the half width at half maximum are shown in table~\ref{tab:HWHM}. Along the $\Gamma \mathrm{K}$ direction the linewidths of the first six branches are far below the experimental resolution (1.3 meV). However, along the $\Gamma \mathrm{M}$ direction, the electron-phonon coupling is responsible of a large broadening of the two soft modes on a wide range of wavevectors. In both directions, the linewidth is almost independent of the smearing of the Fermi surface, driven artificially by the increase of T$_{elec}$, up to T$_{elec}=0.34\,$eV and consistently with previous report \cite[Fig.4(b)]{Weber_NbSe2}.

\begin{table}[h]
\centering
\begin{tabular}{|c||c|c|c|c|c|c|}
\hline
$x \cdot \Gamma$M & $\gamma_1$ &  $\gamma_2$  &
$\gamma_3$ &  $\gamma_4$ &  $\gamma_5$ 
& $\gamma_6$ \\
\hline
0.2 & $<0.01$ & 0.01 & 0.01 & $<0.01$ & 0.04 & 0.03 \\
0.4 & 0.01 & $<0.01$ & 0.06 & 0.05 & 0.36 & 0.35\\
0.6 & 1.17 & 0.89 & 0.04 & 0.03 & 0.18 & 0.12\\
0.8 & 1.52 & 1.04 & 0.07 & 0.06 & 0.19 & 0.15\\
1.0 & 1.36 & 1.14 & 0.03 & 0.08 & 0.12 & 0.09\\
\hline
\hline
 $x \cdot \Gamma$K & $\gamma_1$ &  $\gamma_2$ &
$\gamma_3$ &  $\gamma_4$ &  $\gamma_5$ 
& $\gamma_6$  \\
\hline
1/6 & $<0.01$ & $<0.01$ & $<0.01$ & $<0.01$ & $<0.01$ & $<0.01$ \\

2/6 & $<0.01$ & $<0.01$ & 0.02 & 0.01 & 0.03 & 0.03 \\

3/6 & 0.02 & 0.01 & 0.04 & 0.04 & 0.04 & 0.05 \\

4/6 & 0.02 & 0.02 & 0.06 & 0.06 & 0.05 & 0.06 \\

5/6 & 0.03 & 0.03 & 0.07 & 0.07 & 0.06 & 0.06 \\

6/6 & 0.04 & 0.03 & 0.08 & 0.08 & 0.06 & 0.06 \\
\hline
\end{tabular}
\caption{Phonons linewidths $\gamma$ in meV (Half Width Half
Maximum) of the first 6 phonon modes, calculated ab
initio for several positions along $\Gamma \mathrm{M}$ (top) and  $\Gamma \mathrm{K}$ (bottom). Positions are expressed as a
fraction $x$ of $\Gamma \mathrm{M}$ (resp.$\Gamma$K). At each position $x$, the $\gamma$ are sorted by increasing order of the phonon energy.}
\label{tab:HWHM}
\end{table} 

\begin{figure}
\centering 
\includegraphics[width=8cm]{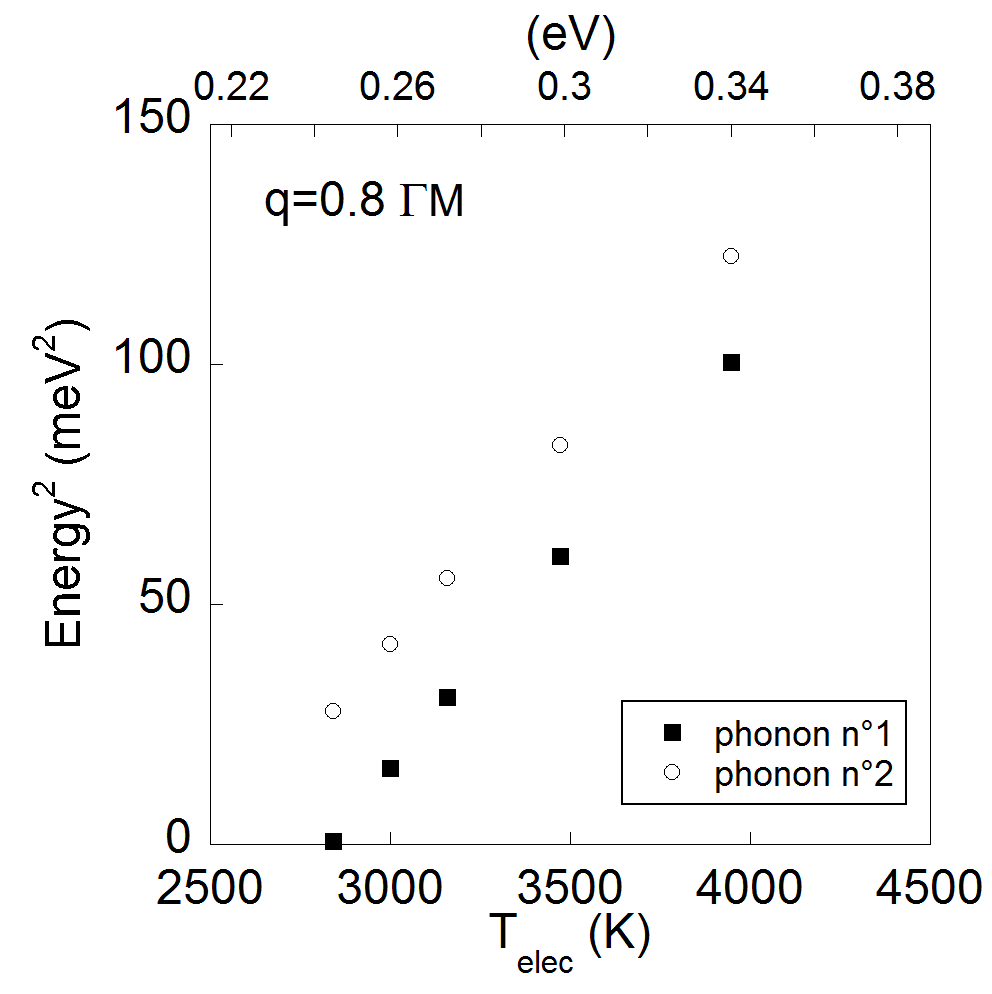}
\caption{Calculated phonon energy squared for the two lowest branches at 0.8$\Gamma \mathrm{M}$ as a function of the smearing of the Fermi surface with T$_{elec}$.}
\label{fig:Telec}
\end{figure}

On the contrary, T$_{elec}$ has a very strong effect on the renormalisation of the phonon energies close to the M point. This is illustrated in Fig.~\ref{fig:Telec}. Yet, a negligible renormalization is observed along the $\Gamma \mathrm{K}$ direction, thus emphasizing the anisotropy of the electron-phonon coupling.

Typical IXS spectra calculated at 0.75$\Gamma M$ for electronic temperatures of $T_{elec}$=0.34 eV and 0.27 eV (including the electron-phonon linewidth, weakly dependent on the chosen electronic temperature, as well as the experimental resolution) are shown in Fig.~\ref{fig:scans_allT} (b) and are in good agreement with the experimental data.
The electronic temperature smears the Fermi surface, and by doing so suppresses the effect of the EPC. This in turn increases the phonon frequency artificially and stabilizes the lattice. However, this increase of $T_{elec}$ is not the correct physical picture since to reproduce changes in spectra in the 0-300 K temperature range, the electronic temperature must be changed by at least $0.1$ eV, and even more in the case of 2H-NbSe$_2$~\cite{Weber_NbSe2} or 1T-TiSe$_2$~\cite{Weber_tise2}. Thus, considering this effect as the origin of the temperature behaviour of the phonon frequencies is incorrect ~\cite{Weber_NbSe2,Weber_tise2}.

On the contrary, the aforementioned anharmonicity, neglected in these first-principle calculations, is likely responsible for these discrepancies.
This was already suggested by Varma and Simons~\cite{Varma_anharmonicCDW}, who demonstrated that anharmonicity can be responsible for a reduction by an order of magnitude of $T_{CDW}$ in 1T-TaS$_2$, hence accounting for the anomalously large ratio of the CDW gap amplitude to $T_{CDW}$ in these compounds.
The strong mean field power law temperature dependence of the soft phonon modes of 2H-NbS$_2$, and the prediction by harmonic DFT calculations of a CDW instability close to the $M$ point, both consistently indicate that anharmonic effects are, in this case, so strong that they actually suppress the CDW and stabilize the crystal structure, leaving however the superconducting properties essentially unaffected~\cite{NasuPRB91}.

In real space, the main components of the soft phonon modes are the vibrations of the Nb atoms in the ({\bf a},{\bf b}) plane, but also the vibrations along the c-axis of a couple of S atoms situated just above and below the plane of Nb atoms. The longitudinal displacement of Nb atoms along $\mathrm{\Gamma M}$ corresponds to a movement toward this couple of S atoms. Interestingly, the distance between these S atoms is unusually short (2.97 \AA$ $,while typical S-S bond are 3.3-3.4 \AA~\cite{Jellinek_1960sulphides}). It therefore suggests that the displacement of Nb atoms is hampered by the great elastic cost of moving apart the S atoms, which would be the microscopic origin of the anharmonicity. The decisive role of the force constant between the S atoms in the same S-Nb-S sandwich was already emphasized in a study based on a tight binding approach~\cite{Motizuki_JPSJ_63_156} and with first principles calculations in TiSe$_2$~\cite{Calandra11_TiSe2}.
In this respect 2H-NbS$_2$ is rather unique as it is on the verge of the CDW transition and is only stabilized by anharmonic effects.

A second important result of our work is that the EPC (responsible for the softening close to the M point) is strongly anisotropic throughout the Brillouin zone. The ab-initio calculation yields a very large phonon linewidth of 1.5 meV at 0.8$\,\Gamma$M, due to the EPC, compared to less than $0.1\,$meV along the $\Gamma$K line. This strong in-plane anisotropy has necessarily important consequences for the superconducting properties of TMD. In this respect it is instructive to compare superconducting gap measurements of the two isoelectronic dichalcogenides 2H-NbSe$_2$ and 2H-NbS$_2$. In the former, there exists two main values of the superconducting gap ~\cite{Garoche1976, HessPRL89, Corcoran1994, Sanchez1995, guillamon134505}. This was initially attributed to the CDW order, presumably competing with the superconducting instability~\cite{Kiss07, Fletcher07, CastroNeto01, Borisenko09}. However a similar observation in the superconducting state of 2H-NbS$_2$ by the same various methods (scanning tunneling spectroscopy, specific heat measurements, $H_{c1}$,  magnetic penetration depth measurement), clearly ruled out this assumption~\cite{guillamonPRL08,Kacmarcik2010,Diener2011,Leroux2012}. As a consequence of the strongly anisotropic softening of the phonons that we report in this study, the EPC, and therefore the pairing strength, are themselves highly anisotropic. This suggests that the two superconducting gap amplitudes observed in 2H-NbS$_2$, and by extension in 2H-NbSe$_2$, rather originate from the anisotropy of the electron-phonon coupling.

\section{Conclusion}

In conclusion, we identified two wide soft phonon modes in 2H-NbS$_2$. They strongly soften along the $\Gamma M$ direction as the temperature is lowered, but their energies always remain finite and never freeze into a static modulation of the lattice. This absence of a CDW is in disagreement with the first-principle harmonic calculations. The mean field power law temperature dependence over the range 300 K-2 K is also an unambiguous evidence of strong anharmonic effects which smear the effect of the EPC. These anharmonic effects are responsible for the absence of CDW in 2H-NbS$_2$, but also for the low $T_{CDW}$ transition temperatures in other TMD.  Thus, 2H-NbS$_2$, which is the lightest compound among the TMD, is a TMD on the verge of a CDW, only hampered by anharmonic effects. These, however, leave the superconducting properties essentially unaffected compared to 2H-NbSe$_2$, which suggests that the relevant parameters for superconductivity in TMD are the anisotropy and strength of the electron-phonon coupling, rather than the nature of the ground state.

\begin{acknowledgments}
Long and fruitful correspondence with L. P. Gor'kov is greatly acknowledged. We thank D. Gambetti for technical assistance. Fruitful discussions with P. Monceau, M. Krisch and X. Blase are also gratefully acknowledged. Experiments were carried out at the European Synchrotron Radiation Facility, on beamlines ID28 (IXS) and ID29 (TDS). Calculations were performed at the IDRIS supercomputing center (proposal number: 091202 ).
\end{acknowledgments}

\bibliography{Leroux_biblio_utf8}

\end{document}